\documentclass[aps,prl,twocolumn,showpacs,preprintnumbers]{revtex4-1}
\usepackage{psfrag,graphicx}
\usepackage{dcolumn}
\usepackage{amsmath,amssymb}
\usepackage{amsfonts}
\usepackage{bm}
\usepackage{amsfonts,amssymb,amsmath}
\usepackage{bbold}
\usepackage{epstopdf}
\usepackage{color}

\usepackage{verbatim}

\DeclareSymbolFont{bbold}{U}{bbold}{m}{n}
\DeclareSymbolFontAlphabet{\mathbbold}{bbold}

\usepackage[caption=false]{subfig}

\begin{document}
	
\title{ Fractional Topological Superconductivity and Parafermion Corner States}
\author{Katharina Laubscher, Daniel Loss, and Jelena Klinovaja}
\affiliation{Department of Physics, University of Basel, Klingelbergstrasse 82, CH-4056 Basel, Switzerland}

\begin{abstract}
We consider a system of weakly coupled Rashba nanowires in the strong spin-orbit interaction (SOI) regime. The nanowires are arranged into two tunnel-coupled layers proximitized by a top and bottom superconductor such that the superconducting phase difference between them is $\pi$. We show that in such a system strong electron-electron interactions can stabilize a helical topological superconducting phase hosting Kramers partners of $\mathbb{Z}_{2m}$ parafermion edge modes, where $m$ is an odd integer determined by the position of the chemical potential. Furthermore, upon turning on a weak in-plane magnetic field, the system is driven into a second-order topological superconducting phase hosting zero-energy $\mathbb{Z}_{2m}$ parafermion bound states localized at two opposite corners of a rectangular sample. As a special case, zero-energy Majorana corner states emerge in the non-interacting limit $m=1$, where the chemical potential is tuned to the SOI energy of the single nanowires.

\end{abstract}

\maketitle

\textit{Introduction.---}%
The search for topological phases of matter has generated an enormous amount of research. Motivated by the discovery and classification of topological insulators (TIs) and topological superconductors (TSCs), the field has been driven by the desire to access phases with increasingly exotic properties. In particular, it has been found that the effects of strong electron-electron interactions can lead to exotic fractionalized phases, which are considered particularly interesting due to their potential use for topological quantum computation. However, only one-dimensional (1D) systems allow for an analytically tractable description of such strong interactions via the bosonization formalism. In order to study strongly interacting systems in more than one dimension, one therefore resorts to the so-called \textit{coupled-wire} approach, where two- or three-dimensional systems are built up from weakly coupled 1D channels, such as nanowires. This approach has proven to be exceptionally fruitful in accessing the fractional counterparts of several well-known topological phases such as fractional quantum Hall states~\cite{Kane2002,Teo2014,Klinovaja2014b}, fractional TIs and TSCs~\cite{Klinovaja2014a,Klinovaja2015,Sagi2014,Sagi2015,Meng2014a,Meng2015b,peter,teo,Santos2015,Neupert2014, par7,Sagi2017}, as well as fractional spin liquids~\cite{Meng2015a,Gorohovsky2015,Huang2016}.

Recently, a lot of interest has been raised by the generalization of conventional TIs/TSCs to so-called \textit{higher-order} TIs/TSCs~\cite{Benalcazar2014,Benalcazar2017,Benalcazar2017b,Song2017,Peng2017,Imhof2017,Geier2018,
Schindler2018,Hsu2018,Ezawa2018b,Ezawa2018c,Zhu2018,Wang2018,Yan2018,Liu2018,Zhang2018,
Wang2018b,Volpez2018,roy1, roy2,Kheirkhah2019}. While a conventional $d$-dimensional TI/TSC exhibits $(d-1)$-dimensional gapless boundary modes, a $d$-dimensional $n$th-order TI/TSC hosts gapless modes at its $(d-n)$-dimensional boundaries. 
While electron-electron interactions have been taken into account in a few cases~\cite{You2018a,You2018b}, the main focus was on non-interacting systems, in particular neglecting the possible existence of exotic fractional phases supporting emergent parafermions. This raises the question whether a coupled-wire approach can be used to extend the class of higher-order topological phases to the {\it fractional} regime. In this work, we show that this is indeed possible and explicitly construct a two-dimensional (2D) fractional second-order TSC exhibiting exotic parafermion corner states.

\begin{figure}[t]
\centering
\includegraphics[scale=1.3]{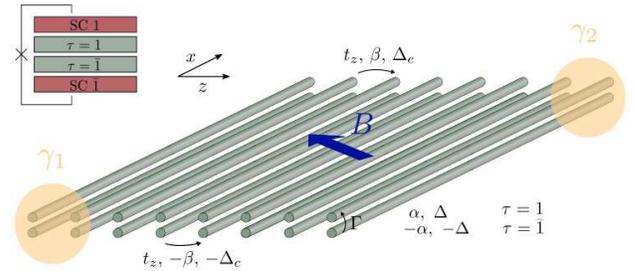}
\caption{The setup consists of two layers of coupled Rashba nanowires where the index $\tau=1$ ($\tau=\bar 1$) denotes the upper (lower) layer. The strength of the Rashba SOI associated with propagation along the $x$ direction is given by $\alpha$ ($-\alpha$) for the upper (lower) layer. Both layers are brought into proximity to an $s$-wave bulk superconductor such that there is a phase difference of $\pi$ between them. In addition, the two layers are strongly coupled by interlayer tunneling of strength $\Gamma$. Neighboring nanowires of the same layer are weakly coupled via a spin-conserving hopping term of strength $t_z$, via a spin-flip hopping term of strength $\beta$ ($-\beta$) associated with Rashba SOI along the $z$ direction, and via a crossed-Andreev superconducting term of strength $\Delta_c$ ($-\Delta_c$), where the last two terms are again of opposite sign for the two layers. Finally, a weak in-plane magnetic field of strength $B$ is applied. We show that in specific regions of parameter space, this system hosts two zero-energy corner states (here denoted by $\gamma_1$ and $\gamma_2$) at two opposite corners of a rectangular sample. In the non-interacting case, these states are Majorana zero modes, whereas strong electron-electron interactions lead to $\mathbb{Z}_{2m}$ parafermion corner states, where $m$ is an odd integer depending on the position of the chemical potential.
}
	\label{fig:wires}
\end{figure}

Our model consists of two layers of coupled Rashba nanowires with proximity-induced superconductivity of a phase difference of $\pi$ between the upper and lower layers, see Fig. \ref{fig:wires}.  In a first step, we show that in the presence of strong electron-electron interactions, such a setup exhibits a helical topological superconducting phase with gapless helical $\mathbb{Z}_{2m}$ parafermion edge modes propagating along the edges. Here, $m$ is  an odd integer determined by the position of the chemical potential $\mu$. In the special case $m=1$, where $\mu$ is tuned to the SOI energy of the single nanowires, Majorana edge modes emerge even in the non-interacting regime. 
At lower densities, the fractional regime $m>1$ emerges in the presence of strong electron-electron interactions as the SOI and Fermi wavevectors get commensurable.

In a second step, we include a small time-reversal breaking perturbation in the form of a weak in-plane magnetic field to gap out the helical edge modes. For a finite rectangular sample, we find $\mathbb{Z}_{2m}$ parafermions localized at two opposite corners of the system depending on the direction of the magnetic field, which places our model in the class of 2D fractional second-order TSCs. Unlike most examples of higher order topological phases, the stability of these corner states does not rely on spatial symmetries but is guaranteed by particle-hole symmetry alone.
Also, the parafermion corner states found here emerge in a spatially uniform 2D system, while in previous studies parafermions have been constructed as bound states localized
at interfaces of non-uniform 2D systems \cite{par1,par2,par3,par4,par4a,par4b,par8a,par8b} or at ends of 1D wires \cite{par5,par6,par5a,par5c,par5b,par5d,par5f,par5e}.

\textit{Model.---}%
We consider two layers of coupled Rashba nanowires proximitized by bulk $s$-wave superconductors, see Fig.~\ref{fig:wires}. Each nanowire of  length $L$ is modeled by a free-particle Hamiltonian
\begin{equation}
\label{eq:kinetic}
H_{0,n}=\sum_{\tau,\sigma}\int dx\ \psi_{n\tau\sigma}^\dagger\Big[-\frac{\hbar^2\partial^2_x}{2m}-\mu
+i\alpha\tau\sigma\partial_x\Big] \psi_{n\tau\sigma}.
\end{equation}
Here, $\psi_{n\tau\sigma}^\dagger(x)$ [$\psi_{n\tau\sigma}(x)$] creates (destroys) an electron at position $x$ in the $n$-th wire in the layer $\tau\in\{1, \bar1\}$ of spin $\sigma\in\{1,\bar1\}$, where we define the spin quantization axis along the SOI direction. The Rashba coefficient $\alpha$ is taken to be of equal magnitude for all nanowires, but of opposite sign for the two layers.
The SOI energy associated with propagation along the nanowire is $E_{so}=\hbar^2k_{so}^2/(2m)$ for $k_{so}=m\alpha/\hbar^2$, and the chemical potential $\mu$ is defined relative to $E_{so}$.
The proximity-induced superconductivity is described by
\begin{equation}
H_{\Delta,n}=\Delta\sum_{\tau}\tau\int dx\,  \psi_{n\tau1}\psi_{n\tau\bar1}+\mathrm{H.c.},
\end{equation}
where we have set the phase difference between the two superconductors to $\pi$. This can, for example, be realized by the Josephson-junction setup shown in the inset of Fig.~\ref{fig:wires}, where the phase difference between the two superconductors is adjusted by controlling the magnetic flux through the superconducting loop~\cite{Ren2018,Fornieri2018}. Alternatively, a thin insulating layer of randomly oriented magnetic impurities~\cite{Schrade2015} could be placed between one of the layers and the corresponding superconductor such that the phase difference of $\pi$ arises due to spin-flip tunneling via the impurities~\cite{Budzin1982,Spivak1991,Ryazanov2001,Dam2006}. 
Furthermore, the two layers are coupled by interlayer tunneling of the form
\begin{equation}
\label{eq:interlayer}
H_{\Gamma,n}=\Gamma \sum_{\tau,\sigma}\int dx\, \psi_{n\tau\sigma}^\dagger\psi_{n\bar\tau\sigma},
\end{equation}
such that the total Hamiltonian describing an effective double nanowire (DNW) composed of two strongly coupled nanowires from different layers is given by $H_n=H_{0,n}+H_{\Delta,n}+H_{\Gamma,n}$. Finally, the DNWs are weakly coupled via a spin-conserving hopping term of strength $t_z$, via a spin-flip hopping term of strength $\beta$ ($-\beta$) associated with Rashba SOI along the $z$ direction as well as via a crossed-Andreev superconducting term of strength $\Delta_c$ ($-\Delta_c$), where the last two terms are again of opposite sign for the two layers. Here, $|t_z|, |\beta|, |\Delta_c|\ll|\Delta|, |\Gamma|$. The interwire Hamiltonian can then be written as
\begin{align}
\label{eq:interwire}
&H_\perp=\sum_{n,\tau,\sigma,\sigma'}\int dx\{ \Delta_c\tau\,\psi_{n\tau\sigma}(i\sigma_y)_{\sigma\sigma'}\psi_{(n+1)\tau\sigma'}/2 \\
&\hspace{20pt}+\psi_{n\tau\sigma}^\dagger[-t_z\delta_{\sigma\sigma'}-i\beta\tau(\sigma_x)_{\sigma\sigma'}/2]\psi_{(n+1)\tau\sigma'}\}+\mathrm{H.c.} \nonumber
\end{align}
 The total Hamiltonian is now given by $H=\sum_{n}H_{n}+H_\perp$, which
in momentum space takes the form $H=\frac{1}{2}\sum_{k_z}\int dk_x\Psi^\dagger_\mathbf{k}\mathcal{H}(\mathbf{k})\Psi_\mathbf{k}$ in the basis $\Psi_\mathbf{k}$=($\psi_{\mathbf{k}1\uparrow}$, $\psi_{\mathbf{k}1\downarrow}$, $\psi_{-\mathbf{k}1\uparrow}^\dagger$, $\psi_{-\mathbf{k}1\downarrow}^\dagger$, $\psi_{\mathbf{k}\bar1\uparrow}$, $\psi_{\mathbf{k}\bar1\downarrow}$, $\psi_{-\mathbf{k}\bar1\uparrow}^\dagger$, $\psi_{-\mathbf{k}\bar1\downarrow}^\dagger$) with
\begin{align}
&\mathcal{H}(\mathbf{k})=\Big[\frac{\hbar^2k_x^2}{2m}-2t_z\mathrm{cos}(k_za_z)-\mu\Big]\eta_z-\alpha k_x\tau_z\sigma_z \label{eq:bulkhamiltonian} \\
&+\beta\mathrm{sin}(k_za_z)\tau_z\sigma_x+\Gamma\tau_x\eta_z+[\Delta+\Delta_c\mathrm{cos}(k_za_z)]\tau_z\eta_y\sigma_y. \nonumber 
\end{align}
Here, $\tau_i$, $\eta_i$, and $\sigma_i$ for $i\in\{x,y,z\}$ are Pauli matrices acting in layer, particle-hole, and spin space, respectively, and $a_z$ is the spacing between neighboring nanowires. The system belongs to the symmetry class DIII~\cite{Ryu2010} with time-reversal (particle-hole) symmetry given by $\mathcal{T}=i\sigma_y\mathcal{K}$ ($\mathcal{P}=\eta_x\mathcal{K}$).

\textit{Helical topological superconducting phase.---}%
We now demonstrate that the system can be brought into a helical topological superconducting phase hosting two counterpropagating $\mathbb{Z}_{2m}$ parafermion edge modes in the presence of strong electron-electron interactions.  For this, we follow the method developed before for fractional TIs \cite{Klinovaja2014a,Klinovaja2015}: First, we solve the DNW Hamiltonian $H_n$ and demonstrate that, due to the interplay between $\Delta$ and $\Gamma$, the elementary excitations are given by gapless $\mathbb{Z}_{2m}$ parafermion modes. We note that, in contrast to Refs.~\cite{Klinovaja2014a,Klinovaja2015}, there are two competing gap-opening mechanisms,
such that when the system is brought close to the critical point $\Gamma\approx\Delta$, again half of the modes are left gapless. Second, we include weak hoppings between DNWs to gap out the parafermion modes in the bulk but leave Kramers pairs of gapless parafermion modes at the edges of the system. Again, if $\beta$ and $\Delta_c$ counterbalance each other, the edge modes propagating along the $x$ axis are perfectly localized at the outermost DNWs. Importantly, the topological phase is robust against deviations from these fine-tuned points, which will, however, lead to increased localization lengths of the edge states.

\begin{figure}[]
	\centering
	\includegraphics[scale=0.8]{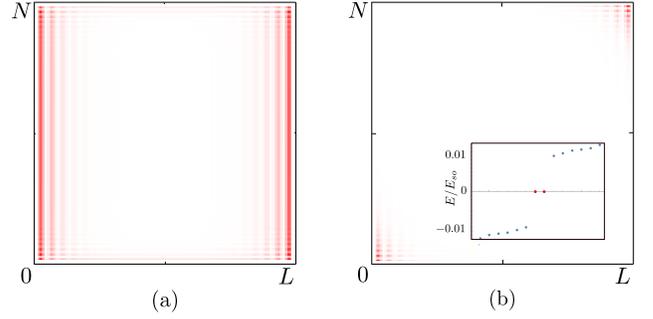}
	\caption{Probability density of low-energy states of $H$ [see 
	Eqs.~(\ref{eq:kinetic})-(\ref{eq:interwire})] obtained numerically.  (a) For $\Delta_Z=0$, the system is a helical TSC with Kramers partners of gapless Majorana modes propagating along the edges. (b) In the presence of a small in-plane magnetic field, $\Delta_Z >0$, we find Majorana bound states localized at two opposite corners of the system. The inset shows the spectrum confirming that these two states (red dots) are indeed at zero energy. The numerical parameters are $N=100$, $\mu=0$, $k_{so}L=85$, $\Gamma/E_{so}\approx0.6$, $\Delta/E_{so}\approx0.55$, $t_z/E_{so}\approx 0.01$, $\beta/E_{so}\approx0.28$, $\Delta_c/E_{so}\approx0.11$, and, in (b), $\Delta_Z/E_{so}\approx0.07$ and $\phi=-\pi/16$. We note that the found topological phases are stable against disorder and do not rely on spatial symmetries.}
	\label{fig:cornerstates}
\end{figure}

For illustrative purposes, we first consider the non-interacting regime with $m=1$ and set $\mu=0$. To treat the DNW Hamiltonian $H_n$, we linearize the spectra of the single nanowires around the Fermi points \cite{Klinovaja2012} as
$\psi_{n\tau\sigma}(x)=R_{n\tau\sigma}(x)e^{ik_F^{1\tau\sigma}x}+L_{n\tau\sigma}(x)e^{ik_F^{\bar 1\tau\sigma}x}$, where $R_{n\tau\sigma}(x)$, $L_{n\tau\sigma}(x)$ vary slowly on the scale of $k_{so}^{-1}$ and the Fermi momenta are given by $k_F^{r\tau\sigma}=(\sigma\tau+r)k_{so}$ ~\cite{mismatch}. We note that upon a change of basis defined by $\bar L_{n\kappa\nu}=(L_{n\kappa\nu}-i\kappa\nu L_{n\bar\kappa\bar\nu})/\sqrt{2}$, $\bar R_{n\kappa\nu}=(R_{n\kappa\nu}-i\kappa\nu R_{n\bar\kappa\bar\nu})/\sqrt{2}$, $H_n$ takes a block-diagonal form, while the structure of the Fermi momenta remains unchanged. For $\Delta\neq0$, the exterior branches $\bar R_{n\kappa\kappa}$ and $\bar L_{n\kappa\bar\kappa}$ are fully gapped by superconductivity, whereas the interior branches $\bar L_{n\kappa\kappa}$ and $\bar R_{n\kappa\bar\kappa}$ have two competing gap-opening mechanisms given by interlayer tunneling and superconductivity. In the following, we thus focus on the interior branches only and tune the system to the critical point $\Delta=\Gamma$. In the new basis, the superconducting and tunneling term take the form $H_{\Gamma,n}=i\Gamma\sum_{\kappa} \int dx \bar R_{n\kappa\bar\kappa}^\dagger \bar L_{n\kappa\kappa}+\mathrm{H.c.}$, $H_{\Delta,n}=\Delta\sum_{\kappa}\int dx \bar R_{n\kappa\bar\kappa}^\dagger\bar L_{n\kappa\kappa}^\dagger +\mathrm{H.c.}$, where the two decoupled sectors labeled by $\kappa$ are related by time-reversal symmetry. Focusing on the first sector (corresponding to $\kappa=1$), we find two gapless counterpropagating Majorana modes per DNW that can be written as
\begin{equation}
\begin{aligned}
\chi_{Ln1}&=(e^{-i\pi/4}\bar L_{n11}+e^{i\pi/4}\bar L_{n11}^\dagger)/\sqrt{2}\, ,\\
\chi_{Rn1}&=(e^{-i\pi/4}\bar R_{n1\bar1}+e^{i\pi/4}\bar R_{n1\bar1}^\dagger)/\sqrt{2}\, .
\end{aligned}\label{eq:zeromodes}
\end{equation}
Next, we add small interwire hopping terms [see Eq.~(\ref{eq:interwire})], where we set $t_z=0$ for simplicity. Focusing on the low-energy sector spanned by the states given in Eq.~(\ref{eq:zeromodes}), $H_\perp$ takes a form similar to a Kitaev chain~\cite{Kitaev2001} of coupled 1D modes,
\begin{equation}
\begin{split}
H_{\perp}&=\frac{i}{2}\sum_{n=1}^{N-1}\int dx[(\beta-\Delta_c)\chi_{R(n+1)1}\chi_{Ln1}\\&\hspace{70pt}+(\beta+\Delta_c)\chi_{L(n+1)1}\chi_{Rn1}],
\end{split}
\end{equation}
where $N$ is the number of DNWs. At the special point $\Delta_c=\beta$, the modes $\chi_{L11}$ and $\chi_{RN1}$ do not enter $H_{\perp}$ and, thus, stay gapless in contrast to all other bulk modes. Obviously, the same is true for their time-reversal partners $\chi_{R1\bar1}$ and $\chi_{LN\bar1}$. Thus, the system is in a helical topological superconducting phase with Kramers partners of gapless Majorana modes propagating along the edges.
Even though this result was derived using a considerable amount of fine-tuning, the topological properties of the system remain qualitatively identical for a broad range of parameters as long as the bulk gap does not close. In particular, our results do not change if a small $t_z$ is included, see Fig.~\ref{fig:cornerstates}(a).

If, on the other hand, the system is infinite along the $z$ axis and finite along the $x$ axis, we apply the standard procedure of matching decaying eigenfunctions~\cite{edge} to find a Kramers pair of gapless Majorana edge modes propagating perpendicular to the DNWs, see the Supplemental Material (SM)~\cite{suppmat} for details and again Fig.~\ref{fig:cornerstates}(a) for a numerical confirmation.

\textit{Fractional helical topological superconducting phase.---}%
Now we focus on the fractional counterpart of the helical superconducting phase discussed above. We tune the chemical potential to a fractional value $\mu/E_{so}=-1+1/m^2$, where $m$ is an odd integer. The new Fermi momenta are now given by $k_F^{r\tau\sigma}=(\tau\sigma+r/m)k_{so}$.
For $m>1$, the interlayer tunneling term given in Eq.~(\ref{eq:interlayer}) no longer conserves momentum and is therefore suppressed. However, for the special values of chemical potential introduced above, momentum-conserving terms can be constructed by including backscattering terms arising from electron-electron interactions~\cite{Teo2014,Giamarchi2004}.
These terms are given by $\tilde H_{\Gamma,n}=i\tilde\Gamma\sum_{\kappa}\int dx(\bar R_{n\kappa\bar\kappa}^\dagger \bar L_{n\kappa\bar\kappa} )^k(\bar R_{n\kappa\bar\kappa}^\dagger\bar L_{n\kappa\kappa})(\bar R_{n\kappa\kappa}^\dagger \bar L_{n\kappa\kappa})^k+\mathrm{H.c.},$ where $k=(m-1)/2$. 
Similarly, we can write down a dressed superconducting term
 $\tilde H_{\Delta,n}=\tilde\Delta\sum_{\kappa,\nu}\kappa\nu\int dx( \bar R_{n\kappa\bar\nu}^\dagger\bar L_{n\kappa\bar\nu})^k(\bar R_{n\kappa\bar\nu}^\dagger\bar L_{n\kappa\nu}^\dagger)(\bar R_{n\kappa\nu}\bar L_{n\kappa\nu}^\dagger )^k+\mathrm{H.c.}$ Here, the coupling constants $\tilde\Gamma\propto\Gamma g_B^{m-1}$ and $\tilde\Delta\propto\Delta g_B^{m-1}$, where $g_B$ is the strength of a single backscattering process caused by electron-electron interactions, are assumed to be large \cite{par6,par5,Sagi2017,Boyanovsky1989}. In order to treat the interacting Hamiltonian analytically, we adapt a bosonized language~\cite{Giamarchi2004}: $\bar R_{n\kappa\nu}(x)=e^{i\phi_{1n\kappa\nu}(x)}$, $\bar L_{n\kappa\nu}(x)=e^{i\phi_{\bar1n\kappa\nu}(x)}$ for bosonic fields $\phi_{ r n\kappa\nu}(x)$ satisfying standard non-local commutation relations.
 The dressed superconducting and tunneling terms can be simplified by introducing new bosonic operators $\eta_{rn\kappa\nu}(x)=\frac{m+1}{2}\phi_{rn\kappa\nu}(x)-\frac{m-1}{2}\phi_{\bar{r}n\kappa\nu}(x)$ obeying the commutation relations
$[\eta_{rn\kappa\nu}(x),\eta_{r'n'\kappa'\nu'}(x')]=i\pi rm\delta_{rr'}\delta_{nn'}\delta_{\kappa\kappa'}\delta_{\nu\nu'}\mathrm{sgn}(x-x')$. The DNW Hamiltonian takes the simple form 
\begin{align}
H_n&=H_{0,n}+2\sum_\kappa\int dx \,[\tilde\Gamma\mathrm{sin}(\eta_{1 n\kappa\bar\kappa}-\eta_{\bar 1 n\kappa\kappa})\label{eq:H_fractional}\\&\hspace{20pt}+\tilde\Delta\mathrm{cos}(\eta_{1 n\kappa\bar\kappa}+\eta_{\bar 1 n\kappa\kappa})+\tilde\Delta\mathrm{cos}(\eta_{1 n\kappa\kappa}+\eta_{\bar 1 n\kappa\bar\kappa})]\nonumber
\end{align}
with $H_{0,n}=\frac{v}{4\pi m}\sum_{r,\kappa,\nu}\int dx (\partial_x\eta_{rn\kappa\nu})^2$, where $v$ is the Fermi velocity and we focus on the special values of Luttinger liquid (LL) parameters $K_{n\kappa\nu}=1/m$. Again, half the modes are fully gapped by superconductivity, while for the other modes superconductivity and interlayer tunneling compete. Introducing conjugate fields $\varphi_{n\kappa}=(\eta_{1 n\kappa\bar\kappa}-\eta_{\bar 1 n\kappa\kappa}-\pi/2)/(2\sqrt m)$, $\theta_{n\kappa}=(\eta_{1 n\kappa\bar\kappa}+\eta_{\bar 1 n\kappa\kappa})/(2\sqrt m)$, the competing part of the above Hamiltonian can be rewritten as
\begin{align}
\label{eq:sinegordon}
&H_n=\sum_{\kappa}\int dx\,\Big\{\frac{v}{2\pi}[(\partial_x\varphi_{n\kappa})^2+(\partial_x\theta_{n\kappa})^2]\\
&\hspace{69pt}+2\tilde\Gamma\mathrm{cos}(2\sqrt m\varphi_{n\kappa})+2\tilde\Delta\mathrm{cos}(2\sqrt m\theta_{n\kappa})\Big\}. \nonumber
\end{align}
For $\tilde\Gamma=\tilde\Delta$, this Hamiltonian corresponds to two time-reversed copies of a well-known self-dual sine-Gordon model~\cite{Lecheminant2002,Zamolodchikov1985}. For $m=1$, we thus expect to find a single gapless Majorana mode per time-reversal sector, which is consistent with our analysis of the non-interacting regime in the previous section. To study the more general case, we start by noting that for our choice of LL parameters, the competing terms have the same scaling dimension, which allows us to explicitly study the properties of the system along the self-dual line. For $m>1$, however, the superconducting and tunneling terms are irrelevant to first order in the renormalization group (RG) analysis, suggesting a flow to a trivial LL fixed point. To resolve this issue, Ref.~\cite{Boyanovsky1989} argued that upon including a third-order term in the RG equations, a multicritical fixed point is encountered, which in our case separates a gapless phase, a phase dominated by superconductivity, and a phase dominated by interlayer tunneling. Such a fixed point has been shown to be described by a $\mathbb{Z}_{2m}$ parafermion theory~\cite{Zamolodchikov1985}, which means in our case that there are two bulk $\mathbb{Z}_{2m}$ parafermion modes related by time-reversal symmetry residing within each DNW.

We now refermionize the above model in order to obtain an explicit expression for specific primary fields~\cite{Zamolodchikov1985} of these parafermion theories. In particular, we define new composite chiral fermion operators
$\bar\psi^{(m)}_{n\kappa\nu}(x)= \bar R^{(m)}_{n\kappa\nu}(x)e^{iq_F^{1\kappa\nu}x}+\bar L^{(m)}_{n\kappa\nu}(x)e^{iq_F^{\bar1\kappa\nu}x}$ with $\bar R^{(m)}_{n\kappa\nu}=e^{i\eta_{1n\kappa\nu}}$, $ \bar L^{(m)}_{n\kappa\nu}= e^{i\eta_{\bar1n\kappa\nu}}$ and 
Fermi momenta $q_F^{r\kappa\nu}=\frac{m+1}{2}k_F^{r\kappa\nu}-\frac{m-1}{2}k_F^{\bar r\kappa\nu}$~\cite{Sagi2014}. The superconducting and tunneling term acting on the interior branches around $q_F=0$ then take the form $H_{\Gamma,n}=i\tilde\Gamma\sum_{\kappa}\int dx   \bar R^{(m)\dagger}_{n\kappa\bar\kappa}\bar L^{(m)}_{n\kappa\kappa}+\mathrm{H.c.}$, $H_{\Delta,n}=\tilde\Delta\sum_{\kappa}\int dx \bar R^{(m)\dagger}_{n\kappa\bar\kappa}\bar L^{(m)\dagger}_{n\kappa\kappa}+\mathrm{H.c.}$, from which we recover the non-interacting case by setting $m=1$.
Forgetting about the underlying model and thinking in terms of the new fermions only, one can perform the same steps as in the non-interacting case to show that the modes
\begin{equation}
\begin{aligned}
\chi^{(m)}_{Ln1}&=(e^{-i\pi/4}\bar L^{(m)}_{n11}+e^{i\pi/4}\bar L^{(m)\dagger}_{n11})/\sqrt{2},\\
\chi^{(m)}_{Rn1}&=(e^{-i\pi/4}\bar R^{(m)}_{n1\bar1}+e^{i\pi/4}\bar R^{(m)\dagger}_{n1\bar1})/\sqrt{2}
\end{aligned}\label{eq:zeromodes_fractional}
\end{equation}
commute with the superconducting and tunneling term, and the same is true for their Kramers partners $\chi^{(m)}_{Ln\bar1}$, $\chi^{(m)}_{Rn\bar1}$. The above solutions satisfy $\chi^{(m)\dagger}_{rn\kappa}=\chi^{(m)}_{rn\kappa}$, which prompts us to identify them as the $\psi_m$ primary fields of the $\mathbb{Z}_{2m}$ parafermion theories 
describing each DNW. Note that these fields are local in terms of electrons, which makes them particularly convenient to handle.

Similar to the non-interacting case, we introduce dressed interwire couplings for $m>1$, which now couple the $\bar R_{n\kappa\nu}^{(m)}$, $\bar L_{n\kappa\nu}^{(m)}$  fields. Assuming that the interwire terms are relevant~\cite{suppmat} and repeating the analysis of the integer case for the modes given in Eq.~(\ref{eq:zeromodes_fractional}), we find that the bulk of the system is fully gapped, while there is a Kramers pair of gapless modes propagating along the edges of a finite sample. These modes correspond to $\psi_m$ primary fields of a $\mathbb{Z}_{2m}$ parafermion theory. However, it is expected \cite{par4} that there are indeed two full $\mathbb{Z}_{2m}$ parafermion theories
 residing at the edges of the system.

\textit{Majorana and parafermion corner states.---}%
We now show that in the presence of a weak in-plane magnetic field, the system enters a second-order topological superconducting phase. Let us start from the (non-interacting) Zeeman Hamiltonian $H_Z=\Delta_Z\sum_{n,\tau,\sigma,\sigma'}\int dx\,\psi_{n\tau\sigma}^\dagger[\mathrm{cos}(\phi)(\sigma_x)_{\sigma\sigma'}+\mathrm{sin}(\phi)(\sigma_z)_{\sigma\sigma'}]\psi_{n\tau\sigma'}$. For $m>1$, momentum-conserving terms are once again constructed by including suitable backscattering processes, such that the dressed term then couples the $\bar R_{n\kappa\nu}^{(m)}$, $\bar L_{n\kappa\nu}^{(m)}$ fields. In the following, we focus on the regime where the magnetic field strength $\tilde\Delta_Z$ is small enough not to modify the bulk structure. However, as time-reversal symmetry is broken, the helical edge modes are gapped out. Assuming that the system size is large such that far away from the corners, all four edges can be treated independently, we calculate the projection of $H_Z$ onto the edge states for all four edges, see the SM~\cite{suppmat}. If we label the edges of a rectangular sample by an index $p=0,...,3$ in counterclockwise order starting from the bottom edge, the projection of the Zeeman Hamiltonian onto the edge $p$ is given by
\begin{equation}
\mathcal{H}_Z^{\mathrm{eff},p}=-\tilde\Delta_Z\mathrm{cos}(\phi+\varphi_p)\gamma_y,
\end{equation}
where we have defined $\varphi_p=p\pi/2$ and $\gamma_y$ is a Pauli matrix acting on the low-energy subspace spanned, in this order, by the low-energy edge mode belonging to the time-reversal sector $\kappa= 1$ and its Kramers partner belonging to the sector $\kappa=\bar 1$. This shows that the mass term changes sign at two corners of the system. Explicitly, the sign change occurs at two diagonally opposite corners of the sample depending on the direction of the magnetic field. For $\phi\in(0,\pi/2)\cup(\pi,3\pi/2)$ [$\phi\in(\pi/2,\pi)\cup(3\pi/2,2\pi)$] the sign change occurs at the top-left and bottom-right (top-right and bottom-left) corners. In the spirit of a Jackiw-Rebbi model~\cite{Jackiw1976}, there are bound states at the corners where the mass term changes sign, which in our case inherit the exotic properties of the propagating modes and thus can be identified as zero-energy $\mathbb{Z}_{2m}$ parafermion corner states. Again, while this result was derived for the local $\psi_m$ fields, we expect that our arguments generalize to the full set of $\mathbb{Z}_{2m}$ primary fields. 
In the non-interacting limit $m=1$, we find zero-energy Majorana corner states, which is verified numerically in Fig.~\ref{fig:cornerstates}(b).

\textit{Conclusions.---}%
We have studied a system consisting of two layers of coupled Rashba nanowires in the presence of interlayer tunneling and proximity-induced superconductivity of a phase difference of $\pi$ between the layers. We have shown that strong electron-electron interactions can stabilize a helical topological superconducting phase exhibiting Kramers partners of gapless $\mathbb{Z}_{2m}$ parafermion edge modes. Upon turning on a small in-plane magnetic field, the system enters a second-order topological superconducting phase hosting exotic zero-energy parafermion bound states at two corners of a rectangular sample depending on the direction of the magnetic field. 

Our analytical approach is limited to the perturbative regime. However, if the system parameters are increased resulting in non-perturbative gaps, the parafermions will still be present as long as the bulk gap is not closed.
Such non-perturbative regimes could be accessed numerically.
Thus, our model provides a proof of principle for the existence of helical fractional TSCs and second-order fractional TSCs and we 
envision it to be a representative of a more general class of systems exhibiting the same parafermionic features.

\acknowledgements
\textit{Acknowledgments.---}%
This work was supported by the Swiss National Science Foundation and NCCR QSIT. This project received funding from the European Union’s Horizon 2020 research and innovation program (ERC Starting Grant, grant agreement No 757725). We acknowledge helpful discussions with Yanick Volpez.

\bibliographystyle{unsrt}

\begin{thebibliography}{}
\bibitem{mismatch} The assumption of equal Fermi momenta for all nanowires can be relaxed to the requirement that the Fermi momenta of only the interior branches are approximately the same. While small deviations from this point do not alter our main conclusions, substantial differences suppress the interlayer tunneling due to momentum mismatch.
\end{thebibliography}

\end{document}


\title{Supplemental Material: Fractional Topological Superconductivity and Parafermion Corner States}
\author{Katharina Laubscher, Daniel Loss, and Jelena Klinovaja}
\affiliation{Department of Physics, University of Basel, Klingelbergstrasse 82, CH-4056 Basel, Switzerland}

\maketitle

\section{Appendix A: Dressed interwire terms}

In this appendix, we explicitly write down the dressed interwire terms coupling the gapless parafermion modes found to reside within each DNW (see the main text). Let us start from the non-interacting case $m=1$. Focusing on the interior branches $\bar L_{n\kappa\kappa}$, $\bar R_{n\kappa\bar\kappa}$ defined in the main text, the interwire term for $t_z=0$ reads 
\begin{equation}
H_\perp=\frac{1}{2}\sum_{n,\kappa}[-i\beta\kappa (\bar L_{n\kappa\kappa}^\dagger \bar R_{(n+1)\kappa\bar\kappa}+\bar R_{n\kappa\bar\kappa}^\dagger \bar L_{(n+1)\kappa\kappa})+\Delta_c (\bar L_{n\kappa\kappa} \bar R_{(n+1)\kappa\bar\kappa}-\bar R_{n\kappa\bar\kappa} \bar L_{(n+1)\kappa\kappa})]+\mathrm{H.c.}
\end{equation}
From the $m=1$ case, momentum-conserving terms can be constructed for $m>1$ by including backscattering processes in a similar way as was discussed for interlayer hopping and superconductivity in the main text. Explicitly, we define the dressed terms as
\begin{equation}
\begin{split}
H_\perp=\frac{1}{2}\sum_{n,\kappa}[&-i\tilde\beta\kappa(\bar L_{n\kappa\kappa}^\dagger \bar R_{n\kappa\kappa})^k(\bar L_{n\kappa\kappa}^\dagger \bar R_{(n+1)\kappa\bar\kappa})(\bar L_{(n+1)\kappa\bar\kappa}^\dagger \bar R_{(n+1)\kappa\bar\kappa})^k
\\&-i\tilde\beta\kappa(\bar R_{n\kappa\bar\kappa}^\dagger \bar L_{n\kappa\bar\kappa})^k(\bar R_{n\kappa\bar\kappa}^\dagger \bar L_{(n+1)\kappa\kappa})(\bar R_{(n+1)\kappa\kappa}^\dagger \bar L_{(n+1)\kappa\kappa})^k\\&+\tilde\Delta_c(\bar L_{n\kappa\kappa} \bar R_{n\kappa\kappa}^\dagger)^k(\bar L_{n\kappa\kappa} \bar R_{(n+1)\kappa\bar\kappa})( \bar L_{(n+1)\kappa\bar\kappa}^\dagger\bar R_{(n+1)\kappa\bar\kappa})^k\\&-\tilde\Delta_c(\bar R_{n\kappa\bar\kappa}\bar L_{n\kappa\bar\kappa}^\dagger )^k(\bar R_{n\kappa\bar\kappa} \bar L_{(n+1)\kappa\kappa})(\bar R_{(n+1)\kappa\kappa}^\dagger\bar L_{(n+1)\kappa\kappa} )^k]+\mathrm{H.c.},
\end{split}
\end{equation}
where again $k=(m-1)/2$. As can easily be checked by changing to bosonic fields, the two competing terms in $H_\perp$ have the same structure as the two competing DNW terms introduced in the main text. Therefore, the arguments for the relevance of $H_\perp$ can be directly adapted from the corresponding arguments for $H_n$. In terms of the composite fermions defined in the main text, the interwire Hamiltonian reads
\begin{equation}
H_\perp=\frac{1}{2}\sum_{n,\kappa}[-i\tilde\beta\kappa (\bar L^{(m)\dagger}_{n\kappa\kappa} \bar R^{(m)}_{(n+1)\kappa\bar\kappa}+\bar R^{(m)\dagger}_{n\kappa\bar\kappa} \bar L^{(m)}_{(n+1)\kappa\kappa})+\tilde\Delta_c (\bar L^{(m)}_{n\kappa\kappa} \bar R^{(m)}_{(n+1)\kappa\bar\kappa}-\bar R^{(m)}_{n\kappa\bar\kappa} \bar L^{(m)}_{(n+1)\kappa\kappa})]+\mathrm{H.c.},
\end{equation}
from where we can repeat the analysis of the non-interacting case.

\section{Appendix B: Edge modes propagating along the $z$ direction}

To confirm the existence of helical edge modes propagating along the $z$ direction, we assume that the system is finite along the $x$ direction and infinite along the $z$ direction and apply the standard procedure of matching decaying eigenfunctions. Once all terms in the Hamiltonian are dressed by suitable backscattering processes (see the main text as well as Appendix~A), it is convenient to work directly in terms of the composite fermions $\bar \psi_{n\kappa\nu}^{(m)}$ for general $m$. Changing to momentum space along the $z$ axis, we write the problem in terms of the Fourier-transformed fields $\bar \psi_{k_z\kappa\nu}^{(m)}$ and linearize the spectrum around the Fermi points~\cite{linear},
\begin{equation}
\begin{aligned}
\bar\psi_{k_z11}^{(m)}(x)&=\bar R_{k_z11}^{(m)}(x)e^{2ik_{so}x}+\bar L_{k_z11}^{(m)}(x),\\
\bar\psi_{k_z1\bar1}^{(m)}(x)&=\bar R_{k_z1\bar1}^{(m)}(x)+\bar L_{k_z1\bar1}^{(m)}(x)e^{-2ik_{so}x},\\
\bar\psi_{k_z\bar 11}^{(m)}(x)&=\bar R_{k_z\bar 11}^{(m)}(x)+\bar L_{k_z\bar 11}^{(m)}(x)e^{-2ik_{so}x},\\
\bar\psi_{k_z\bar 1\bar1}^{(m)}(x)&=\bar R_{k_z\bar 1\bar1}^{(m)}(x)e^{2ik_{so}x}+\bar L_{k_z\bar 1\bar1}^{(m)}(x).
\end{aligned}
\label{eq:lowenergy}
\end{equation}
Here, $\bar R^{(m)}_{k_z\kappa\nu}(x)$ [$\bar L^{(m)}_{k_z\kappa\nu}(x)$] are again slowly varying right-moving (left-moving) fields. The total Hamiltonian separates into a part corresponding to the exterior branches and a part corresponding to the interior branches~\cite{stano} given by
\begin{equation}
\mathcal{H}_\mathrm{int}=i\hbar v\kappa_z\nu_z\partial_x+\tilde\beta\mathrm{sin}(k_za_z)\kappa_z\nu_x+\tilde\Gamma\kappa_z\nu_y+[\tilde\Delta+\mathrm{cos}(k_za_z)\tilde\Delta_c]\kappa_z\eta_y\nu_y
\end{equation}
in the basis $\bar\Psi^{(m)}_{\mathrm{int}}=(\bar L_{k_z11}^{(m)},\bar R_{k_z1\bar1}^{(m)},\bar L_{-k_z11}^{(m)\dagger},\bar R_{-k_z1\bar1}^{(m)\dagger},\bar R_{k_z\bar11}^{(m)},\bar L_{k_z\bar1\bar1}^{(m)},\bar R_{-k_z\bar11}^{(m)\dagger},\bar L_{-k_z\bar1\bar1}^{(m)\dagger})$ for the interior branches and
\begin{equation}
\mathcal{H}_\mathrm{ext}=-i\hbar v\kappa_z\nu_z\partial_x+[\tilde\Delta+\mathrm{cos}(k_za_z)\tilde\Delta_c]\kappa_z\eta_y\nu_y
\end{equation}
in the basis $\bar\Psi^{(m)}_{\mathrm{ext}}=(\bar R_{k_z11}^{(m)},\bar L_{k_z1\bar1}^{(m)},\bar R_{-k_z11}^{(m)\dagger},\bar L_{-k_z1\bar1}^{(m)\dagger},\bar L_{k_z\bar11}^{(m)},\bar R_{k_z\bar1\bar1}^{(m)},\bar L_{-k_z\bar11}^{(m)\dagger},\bar R_{-k_z\bar1\bar1}^{(m)\dagger})$ for the exterior branches. Here, $\kappa_i$ and $\nu_i$ for $i\in\{x,y,z\}$ are Pauli matrices, and the two sectors labeled by $\kappa$ are related by time-reversal symmetry.
As in the main text, we focus on the regime $\tilde\Gamma\approx\tilde\Delta$ and $\tilde\beta,\tilde\Delta_c\ll\tilde\Gamma,\tilde\Delta$, and assume that all terms in the Hamiltonian are RG-relevant for all $m$. For $\tilde\Delta,\tilde\Delta_c>0$, we then find Kramers pairs of zero-energy solutions at $k_za_z=\pi$ which are exponentially localized to the system edges at $x=0,L$. To demonstrate this, we consider the Hamiltonians
\begin{align}
\mathcal{H}_\mathrm{int}^{(a_zk_z=\pi)}&=i\hbar v\kappa_z\nu_z\partial_x+\tilde\Gamma\kappa_z\nu_y+(\tilde\Delta-\tilde\Delta_c)\kappa_z\eta_y\nu_y,\label{eq:Hzdirection_int}\\
\mathcal{H}_\mathrm{ext}^{(a_zk_z=\pi)}&=-i\hbar v\kappa_z\nu_z\partial_x+(\tilde\Delta-\tilde\Delta_c)\kappa_z\eta_y\nu_y,
\end{align}
and look for exponentially decaying zero-energy eigenfunctions. In particular, we find four solutions corresponding to interior modes,
\begin{equation}
\begin{split}
\phi_1^\mathrm{int}(x)&=(-i,i,-1,1,0,0,0,0)^T\,e^{-x/\xi_1},\\
\phi_2^\mathrm{int}(x)&=(0,0,0,0,-i,i,-1,1)^T\,e^{-x/\xi_1},\\
\phi_3^\mathrm{int}(x)&=(i,-i,-1,1,0,0,0,0)^T\,e^{-x/\xi_1'},\\
\phi_4^\mathrm{int}(x)&=(0,0,0,0,i,-i,-1,1)^T\,e^{-x/\xi_1'},
\end{split}
\end{equation}
where $\xi_1=\hbar v/(\tilde\Gamma-\tilde\Delta+\tilde\Delta_c)$ and $\xi_1'=\hbar v/(\tilde\Gamma+\tilde\Delta-\tilde\Delta_c)$. Similarly, we find four solutions corresponding to exterior modes,
\begin{equation}
\begin{split}
\phi_1^\mathrm{ext}(x)&=(-i,0,0,1,0,0,0,0)^T\,e^{-x/\xi_2},\\
\phi_2^\mathrm{ext}(x)&=(0,-i,1,0,0,0,0,0)^T\,e^{-x/\xi_2},\\
\phi_3^\mathrm{ext}(x)&=(0,0,0,0,-i,0,0,1)^T\,e^{-x/\xi_2},\\
\phi_4^\mathrm{ext}(x)&= (0,0,0,0,0,-i,1,0)^T\,e^{-x/\xi_2},
\end{split}
\end{equation}
with $\xi_2=\hbar v/(\tilde\Delta-\tilde\Delta_c)$. In the regime $\tilde\Gamma, \tilde\Delta>\tilde\Delta_c>0$, $|\tilde\Gamma-\tilde\Delta|<\tilde\Delta_c$, the above solutions are exponentially localized to the left edge of the system at $x=0$. By reinstating the oscillating factors $e^{\pm 2ik_{so}x}$ and imposing vanishing boundary conditions $\Phi^{(m)}_{\pm}(x=0)=0$, we find that one solution at the left edge of the system is given by
\begin{equation}
\Phi_+^{(m)}(x)=(e^{i\pi/4}f,-e^{i\pi/4}f^*,e^{-i\pi/4}f^*,-e^{-i\pi/4}f,0,0,0,0)^T,
\end{equation}
where $f(x)=e^{-2ik_{so}x}e^{-x/\xi_2}-e^{-x/\xi_1}$ and where we omitted the normalization factor.
Its Kramers partner can be obtained by time-reversal symmetry as $\Phi^{(m)}_-=-\mathcal{\bar T}\Phi^{(m)}_+$, and both solutions satisfy $\mathcal{\bar P}\Phi^{(m)}_\pm=\Phi^{(m)}_\pm$. Here, $\mathcal{\bar T}$  and $\mathcal{\bar P}$ are the representations of time-reversal and particle-hole symmetry in the new basis $\bar\Psi^{(m)}$,  respectively, which are explicitly given by $\mathcal{\bar T}=\kappa_y\eta_z\mathcal{K}$ and $\mathcal{\bar P}=\eta_x\mathcal{K}$.

Together with the solutions along the $x$ direction obtained in the main text, we can now combine the results for both semi-infinite geometries to conclude that there is a single pair of counterpropagating Majorana edge modes for a system which is large but finite both along the $x$ and $z$ direction.

\section{Appendix C: Effective edge Hamiltonian}

In order to calculate the effective low-energy Hamiltonian describing the gap opened in the spectrum of edge states [see Eq.~(11) in the main text], we start by expressing the Zeeman part of the Hamiltonian in terms of the new basis $\bar\Psi^{(1)}$ and include suitable backscattering processes for $m>1$. We obtain the Hamiltonian
\begin{equation}
\mathcal{H}_Z=\tilde\Delta_Z[\mathrm{cos}(\phi)\kappa_y\nu_z-\mathrm{sin}(\phi)\kappa_y\nu_x]
\end{equation}
in the basis $\bar\Psi^{(m)}$. Again, we assume that this term is relevant in the RG sense for all $m$. Let us first consider the edges aligned along the $x$ direction. If the system is assumed to be infinite along the $x$ direction, we find two Kramers partners of zero-energy wave functions at $k_x=0$, which we label as $\Phi_{0,\pm}$ for the bottom edge ($p=0$) and $\Phi_{2,\pm}$ for the top edge ($p=2$) of the system in correspondence with the labeling of the edges used in the main text. We note that the DNW Hamiltonian exhibits an additional symmetry corresponding to the operator $\mathcal{O}_1=\eta_y\nu_z$, which anticommutes with both interlayer tunneling as well as superconductivity. Furthermore, $\mathcal{O}_1$ commutes with the particle-hole symmetry operator $\mathcal{\bar P}$. For the edge states $\Phi_{0,+}^{(m)}$ and $\Phi_{0,-}^{(m)}=-\mathcal{\bar T}\Phi_{0,+}^{(m)}$, we find $\mathcal{O}_1\Phi_{0,\pm}^{(m)}=-\Phi_{0,\pm}^{(m)}$. We thus arrive at
\begin{equation}
\langle\Phi^{(m)}_{0,+}|\kappa_y\nu_x|\Phi^{(m)}_{0,-}\rangle=\langle\Phi^{(m)}_{0,+}|\mathcal{O}_1\kappa_y\nu_x\mathcal{O}_1|\Phi^{(m)}_{0,-}\rangle=-\langle\Phi^{(m)}_{0,+}|\kappa_y\nu_x|\Phi^{(m)}_{0,-}\rangle,
\end{equation}
and therefore $\langle\Phi_{0,+}^{(m)}|\kappa_y\nu_x|\Phi_{0,-}^{(m)}\rangle=0$. Here, we used that $\{\mathcal{O}_1,\kappa_y\nu_x\}=0$. On the other hand, we have
\begin{equation}
\langle\Phi_{0,+}^{(m)}|\kappa_y\nu_z|\Phi_{0,-}^{(m)}\rangle=\langle\Phi_{0,+}^{(m)}|\kappa_y\nu_z(-\kappa_y\eta_z\mathcal{K})(\eta_x\mathcal{K})|\Phi_{0,+}^{(m)}\rangle=-i\langle\Phi_{0,+}^{(m)}|\mathcal{O}_1|\Phi_{0,+}^{(m)}\rangle=i.
\end{equation}
Hence, we find
\begin{equation}
\mathcal{H}_Z^\mathrm{eff, p=0}=-\tilde\Delta_Z\mathrm{cos}(\phi)\gamma_y,
\end{equation}
where $\gamma_y$ is a Pauli matrix acting on the low-energy subspace spanned by $\Phi_{0,\pm}^{(m)}$.
In order to calculate the effective Hamiltonian for the top edge, we note that our system is invariant under rotation around the $y$ axis by an angle $\pi$, which leads to
\begin{equation}
\mathcal{H}_{Z}^{\mathrm{eff,p=2}}=-\mathcal{H}_{Z}^{\mathrm{eff,p=0}}=\tilde\Delta_Z\mathrm{cos}(\phi)\gamma_y.
\end{equation}
We now treat the edges along the $z$ direction, where we use the properly normalized zero-energy wave functions found in Appendix B and label them as $\Phi_{1,\pm}^{(m)}$ for the right edge and $\Phi_{3,\pm}^{(m)}$ for the left edge. Again, we can use certain symmetries of the system to calculate the above matrix elements. In particular, the operator $\mathcal{O}_2=\eta_y\nu_x$ anticommutes with the Hamiltonian at $k_za_z=\pi$. At the same time, we have $[\mathcal{O}_2,\mathcal{\bar P}]=0$. An explicit calculation yields $\mathcal{O}_2\Phi_{3,\pm}^{(m)}(x)= \Phi_{3,\pm}^{(m)}(x)$. Again, we can use the fact that $\{\mathcal{O}_2,\kappa_y\nu_z\}=0$ to argue that $\langle\Phi^{(m)}_{3,+}|\kappa_y\nu_z|\Phi^{(m)}_{3,-}\rangle=0$. 
On the other hand, we find
\begin{equation}
\langle\Phi^{(m)}_{3,+}|\kappa_y\nu_x|\Phi^{(m)}_{3,-}\rangle=\langle\Phi^{(m)}_{3,+}|\kappa_y\nu_x(-\kappa_y\eta_z\mathcal{K})(\eta_x\mathcal{K})|\Phi^{(m)}_{3,+}\rangle=-i\langle\Phi^{(m)}_{3,+}|\mathcal{O}_2|\Phi^{(m)}_{3,+}\rangle=-i.
\end{equation}
Therefore, we arrive at
\begin{equation}
\mathcal{H}_Z^\mathrm{eff, p=3}=-\tilde\Delta_Z\mathrm{sin}(\phi)\gamma_y.
\end{equation}
Again, the effective Hamiltonian for the right edge can be obtained by exploiting the two-fold rotation symmetry of the system, which gives us
\begin{equation}
\mathcal{H}_{Z}^{\mathrm{eff,p=1}}=-\mathcal{H}_{Z}^{\mathrm{eff,p=3}}=\tilde\Delta_Z\mathrm{sin}(\phi)\gamma_y.
\end{equation}
Combining the above results, we arrive at the effective Hamiltonian given in Eq.~(11) of the main text. Following Ref.~\cite{Jackiw1976}, we conclude that there exist zero-energy bound states at the corners where the mass term changes sign. Importantly, we note that this argument is independent of any gauge choice. If one would naively multiply an arbitrary phase factor to a solution on a particular edge, the time-reversal relation between the two Kramers partners at this edge changes, while the corresponding relations stay unmodified for all other edges, which would then contradict the idea of the solutions being connected to form a single set of counterpropagating edge modes.

\bibliographystyle{unsrt}